\documentclass[journal,onecolumn, 12pt]{IEEEtran}
\usepackage{cite}
\usepackage[table]{xcolor}
\usepackage{amsmath,amssymb,amsfonts}
\usepackage{algorithmic}
\usepackage{graphicx}
\usepackage{subcaption}
\usepackage{array}
\usepackage{arydshln}
\usepackage{balance}
\usepackage{makecell}
\usepackage{booktabs}
\usepackage[]{footmisc}
\usepackage[margin=1in]{geometry}

\ifCLASSINFOpdf
\else
\fi
\begin{document}
%
\title{Nanoscale Communication: State-of-Art and Recent Advances}
%
%
%

\title{Nanoscale Communication: State-of-Art and Recent Advances}
\author{\IEEEauthorblockN{\textbf{Hadeel Mohammad}$^{*}$, \textbf{Raed M. Shubair}$^\ddag$}\\
\IEEEauthorblockA{$^*$Electrical and Computer Engineering, University of Toronto, Canada \\
$^\ddag$Research Laboratory of Electronics, Massachusetts Institute of Technology, USA\\ 
Emails: hadeel@tornoto.edu ; rshubair@mit.edu}
}
\maketitle

\begin{abstract}
The engineering community is witnessing a new frontier in the communication industry. Among others, the tools provided by nanotechnologies enable the development of novel nanosensors and nanomachines. On the one hand, nanosensors are capable of detecting events with unprecedented accuracy. On the other hand, nanomachines are envisioned to accomplish tasks ranging from computing and data storing to sensing and actuation. Recently, in vivo wireless nanosensor networks (iWNSNs) have been presented to provide fast and accurate disease diagnosis and treatment. These networks are capable of operating inside the human body in real time and will be of great benefit for medical monitoring and medical implant communication. Despite the fact that nanodevice technology has been witnessing great advancements, enabling the communication among nanomachines is still a major challenge.
\end{abstract}


%
\IEEEpeerreviewmaketitle


\section{Chapter Overview}
The last two decades witnessed an exponential growth and tremendous developments in wireless technologies and systems, and their associated applications, such as those reported in \cite{omar2016uwb,elayan2017terahertz,elayan2017wireless,elayan2016channel,elayan2016vivo,elayan2017bio,elayan2018end,elayan2017photothermal,elayan2017multi,elayan2018vivo,alnabooda2017terahertz,khan2017ultra,khan2017compact,shubair2015novel,shubair2015vivo,alayyan2009mmse,elayan2018towards,kulaib2015improved,elayan2018stochastic,elayan2018graphene,elayan2018characterising,elayan2018terahertz,alharbi2018flexible,kiourti2017implantable,khan2017second,ibrahim2017compact,khan2016properties,khan2016pattern,elayan2016revolutionizing,el2016design,bazazeh2016biomarker,bazazeh2016comparative,albreiki2016coding,shubair2015survey,elsalamouny2015novel,nwalozie2013simple,che2008propagation,bakhar2009eigen,khan2018compact,khan2016compact,al2005direction}. In recent decades, the technological advances in novel materials have enabled a new generation of increasingly smaller electronics, which are fundamental tools for the future development of components such as processors, batteries, sensors and actuators. The downsizing of electronics to the scale of a few nanometers opens the analysis of different, unforeseen essential parameters and magnitudes, such as hormone levels, disease detection, control of bio-implants in human or animal bodies as well as air pollution measurements in the atmosphere, among others.

In this chapter, the concept of nanonetworks is going to be discussed along with an extensive explanation of the nanomachine hardware architecture. In addition, a plethora of potential applications in fields such as biomedicine, environmental sciences and industry are going to be addressed. The chapter concludes by presenting current developments of nanoscale devices which aim to provide additional capacities for attaining novel technological solutions.  
\section{Nanonetworks}
\begin{figure}[h!]
\centering
\includegraphics[width=0.8\textwidth]
{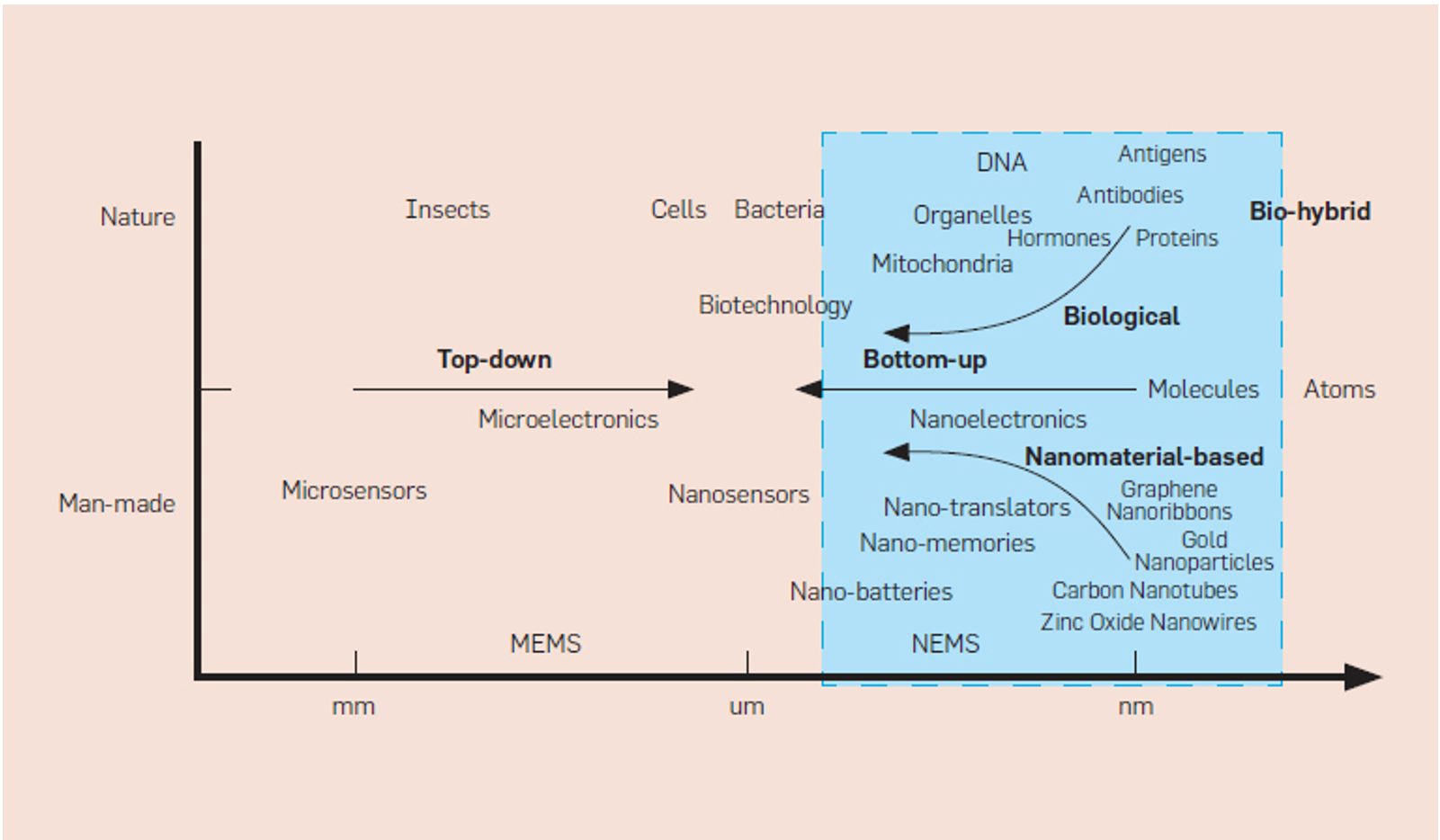}
\caption{Illustration of the concept behind nanomachine design \cite{jornet2013fundamentals}.}
\label{fig:topdown}
\end{figure}

 Nanotechnology is providing a new set of tools to the engineering community to create nanoscale components that are able to perform simple tasks, such as computing, data storing, sensing, and actuation. Integrating several of these nanocomponents into a single device just a few cubic micrometers in size will enable the development of more advanced nanodevices and will further allow these devices to achieve complex tasks in a distributed manner \cite{akyildiz2008nanonetworks}. The resulting nanonetworks will permit unique applications in the biomedical, industrial, and military fields, such as advanced health monitoring systems, nanosensor networks for biological and chemical attack prevention as well as wireless network on chip systems for very large multicore computing architecture \cite{akyildiz2010electromagnetic}. Fig \ref{fig:topdown} illustrates the concept behind nanomachine design in which it could be inspired by following either a top-down or a bottom-up approach, which are two major means commonly utilized in product manufacturing.  

Several communication paradigms can be used in nanonetworks depending on the technology deployed to synthesize the nanomachine and the targeted application. In this chapter, the focus will be on nanoelectromagnetic communication which is defined as the transmission and reception of electromagnetic radiation from components based on novel nanomaterials \cite{rutherglen2009nanoelectromagnetics}. From a communication perspective, the unique properties observed in novel nanomaterials will decide on the specific bandwidths for emission of electromagnetic radiation, the time lag of the emission or the magnitude of the emitted power for a given input energy \cite{akyildiz2010propagation}.

\section{Nanomachine Hardware Architecture}
\begin{figure}[h!]
\centering
\includegraphics[width=0.8\textwidth]
{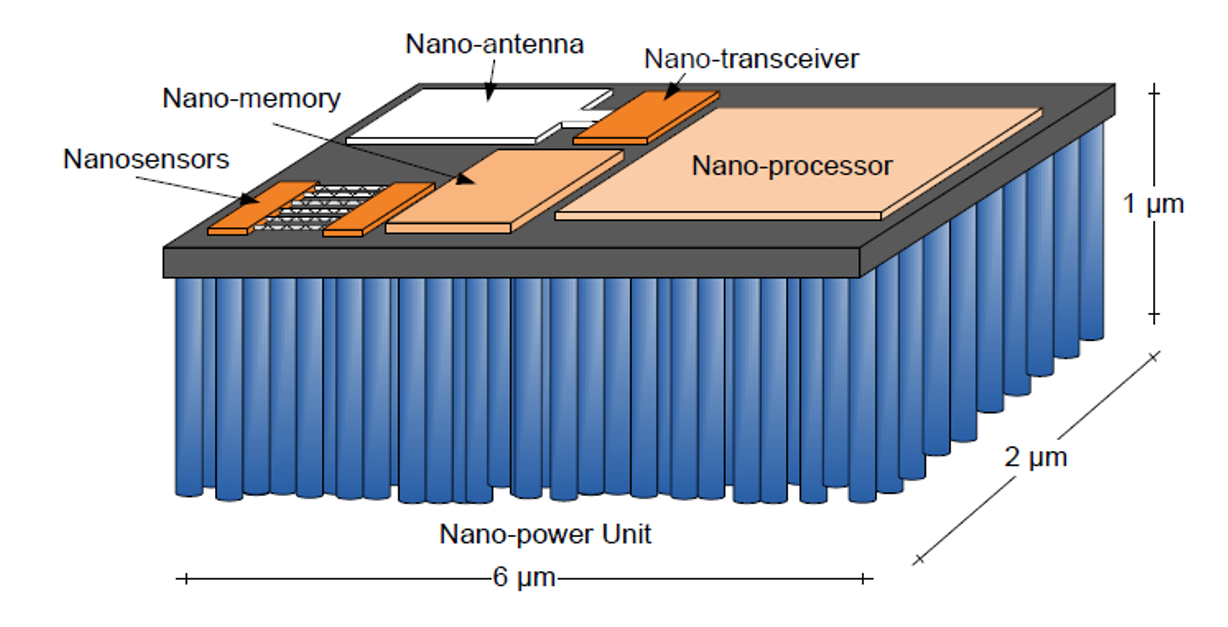}
\caption{Nanomachine hardware architecture \cite{jornet2013fundamentals}.}
\label{fig:nanonetwork}
\end{figure}

There are numerous challenges in the development of autonomous nanomachines. In Fig. \ref{fig:nanonetwork}, a conceptual nanomachine architecture is shown. To the best of our knowledge, fully functional
nanomachines have not been built to date. However, several solutions for each nanocomponent have been prototyped and tested as follows.
\begin{itemize}
\item 
Processing Unit: Nanoprocessors are being enabled by the development of tinier Field Effect Transistors (FET)  in different forms. The smallest transistor that has been experimentally tested to date is based on a thin graphene strip made of just 10 by 1 carbon atoms \cite{ponomarenko2008chaotic}. These transistors are not only smaller, but also able to operate at higher frequencies. The complexity of the operations that a nanoprocessor will be able to handle directly depends on the number of integrated transistors in the chip, thus, on total size.
Table \ref{table:1} reviews the related transistor manufacturing technologies, comparing their scalability, power consumption and fabrication process.  
\begin{table}[h!]
\captionsetup{skip=0pt,font=footnotesize}
\caption{Comparison of Nanoprocessor Technologies \cite{canovas2016conceptual} }\begin{center}
\centering
\label{table:1}
\small
\begin{tabular}{|p{85pt}| p{70pt}|p{120pt}| p{135pt}| p{45pt}|}\hline
\begin{footnotesize}  \begin{center}\textbf{Transistor Technology}\end{center}\end{footnotesize} & \begin{footnotesize}\begin{center}\textbf{Minimum Transistor Size}\end{center} \end{footnotesize} & \begin{footnotesize}\begin{center}\textbf{Advantages}\end{center} \end{footnotesize} & \begin{footnotesize}\begin{center}\textbf{Disadvantages}\end{center}\end{footnotesize}& \begin{footnotesize}\begin{center}\textbf{Feasibility}\end{center}\end{footnotesize}\\\hline

\raisebox{10pt}

\begin{footnotesize} Silicon\end{footnotesize} & \begin{footnotesize} 14 nm \end{footnotesize} & \begin{footnotesize} Mature Technology

Low-cost manufacturing\end{footnotesize}&  \begin{footnotesize}Scalability concerns\end{footnotesize}& \begin{footnotesize}Yes\end{footnotesize}   \\\hline
\raisebox{10pt}

\begin{footnotesize}Silicon Germanium (SiGe) \end{footnotesize}  &\begin{footnotesize} 7 nm \end{footnotesize} &\begin{footnotesize} Good scalability

Low-cost manufacturing

Ultra-low power consumption \end{footnotesize} & \begin{footnotesize}Experimental technology\end{footnotesize}&\begin{footnotesize}Yes\end{footnotesize}\\\hline
\raisebox{10pt}

\begin{footnotesize}Carbon Nanotube (CNT) \end{footnotesize}  &\begin{footnotesize} Sub 20 nm\end{footnotesize}& \begin{footnotesize} Great scalability 

Ultra-low power consumption
High speed
\end{footnotesize} & \begin{footnotesize}Experimental technology

Difficult manufacturing\end{footnotesize} & \begin{footnotesize}Yes\end{footnotesize} \\\hline
\raisebox{10pt}

\begin{footnotesize} Atomic \end{footnotesize}  &\begin{footnotesize} One atom thick\end{footnotesize}&\begin{footnotesize} Ultra-small size\end{footnotesize} & \begin{footnotesize} Operation under strict laboratory conditions\end{footnotesize}  &\begin{footnotesize}Not yet\end{footnotesize} \\\hline

\end{tabular}
\end{center}
\end{table}
\item Data Storage Unit: The storage capacity of an electronic device is an important aspect since the amount and complexity of the stored programming codes rely directly on the available memory. This has an impact on most nanodevice functionalities, as for instance, the communication protocol stack. In this sense, many of its configuration parameters (such as device ID length, packet size, number of bits for error detection, etc.) intrinsically depend on the available memory. Nanomaterials and new manufacturing processes are actually enabling the development of single-atom nanomemories, in which the storage of one bit of information requires only one atom \cite{bennewitz2002atomic}. For example, in a magnetic memory \cite{parkin2008magnetic}, atoms are placed over a surface by means of magnetic forces. While these memories are not ready yet for nanomachines, they serve as a starting point. The total amount of information storable in a nanomemory will ultimately depend on its dimensions.

\item Power Unit: Powering nanomachines require new types of nanobatteries \cite{ji2011multilayer} as well as nanoscale energy harvesting systems \cite{wang2008towards}. One of the most promising techniques relies on the piezoelectric effect seen in zinc oxide nanowires, which are used to convert vibrational energy into electricity. This energy can then be stored in a nanobattery and dynamically consumed by the device. The rate at which energy is harvested and the total energy that can be stored in a nanodevice depends ultimately on the device size. Recall that as a general design requirement, our nanodevice size should be similar to the size of a blood cell. This tiny size makes it unfeasible to manipulate it to replace a depleted battery. Thus, to guarantee an appropriate power level to feed the nanodevice efficiently, we consider  two solutions: (i) harvesting the energy from the environment (denoted as self-powered nanodevice); and (ii) wireless energy induced from an external power source \cite{wang2006piezoelectric}. Table \ref{table:3}  compares among existing storing technologies considering their main features.  
\begin{table}[h!]
\captionsetup{skip=0pt,font=footnotesize}
\caption{Comparison Among Storing Technologies \cite{canovas2016conceptual} }\begin{center}
\centering
\label{table:3}
\small
\begin{tabular}{|p{70pt}|p{100pt}| p{95pt}| p{50pt}|}\hline
\begin{footnotesize}  \begin{center}\textbf{Storage Technology}\end{center}\end{footnotesize}  & \begin{footnotesize}\begin{center}\textbf{Advantages}\end{center} \end{footnotesize} & \begin{footnotesize}\begin{center}\textbf{Disadvantages}\end{center}\end{footnotesize}& \begin{footnotesize}\begin{center}\textbf{Feasibility}\end{center}\end{footnotesize} \\\hline

\raisebox{10pt}

\begin{footnotesize}Batteries\end{footnotesize} & \begin{footnotesize} High energy density \end{footnotesize} & \begin{footnotesize} High degradation

Mechanical properties

Use of toxic materials\end{footnotesize}&  \begin{footnotesize}Not clear\end{footnotesize}   \\\hline
\raisebox{10pt}

\begin{footnotesize}Supercapacitors\end{footnotesize}&\begin{footnotesize} High capacitance

Ultra low degradation

Mechanical properties 

Non-toxic materials\end{footnotesize} &\begin{footnotesize} 
Low energy density
 \end{footnotesize} & \begin{footnotesize}Yes\end{footnotesize}\\\hline
\end{tabular}
\end{center}
\end{table}

\item Sensing Unit: Physical, chemical and biological nanosensors have been developed by using graphene and other nanomaterials \cite{hierold2007nano}. A nanosensor is not just a tiny sensor, but a device that makes use of the novel properties of nanomaterials to identify and measure new types of events in the nanoscale, such as the physical characteristics of structures just a few nanometers in size, chemical compounds in concentrations as low as one part per billion, or the presence of biological agents such as virus, bacteria or cancerous cells. Their accuracy and timeliness is much higher than those of existing sensors.
\item Communication Unit: The miniaturization of an antenna to meet the size constraints of nanomachines would impose the use of very high frequencies. This would limit the feasibility of electromagnetic nanonetworks due to the energy limitations of nanomachines. Nanomaterials can be used to develop new types of nanoantennas as well as nanotransceivers, which can operate at much lower frequencies than miniature metallic antennas. However, these introduce many challenges for the realization of communication in nanonetworks.
\end{itemize}

\section{Applications of Nanonetworks}
The most common application areas of nanonetworks will be biology, medicine, chemistry, environmental science as well as the development of military, industrial and consumer goods \cite{akyildiz2010electromagnetic}. In the area of biomedicine, applications such as health monitoring systems that probe the amount of sodium, glucose and other ions in the blood or drug delivery systems that distribute drugs to special parts of the body with controlled doses are envisioned. Plant monitoring systems as well as plague defeating systems are preliminary environmental applications. As for industrial applications, nanosensors could be used in developing new touch surfaces or haptic interfaces. They could be also used to design equipments required for augmented reality or game applications. The future may reveal some new applications that are now not even envisioned.
\subsection{Biomedical Applications}
\begin{figure}[h!]
\centering
\includegraphics[width=0.65\textwidth]
{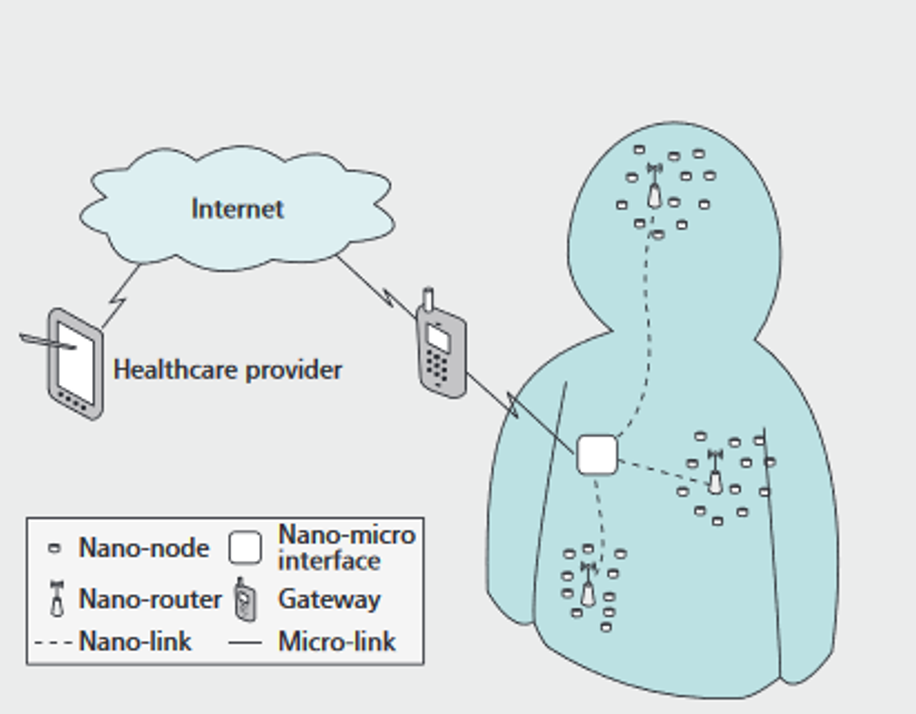}
\caption{ Intrabody nanonetworks for healthcare applications \cite{akyildiz2010internet}.}
\label{fig:healthcare}
\end{figure}

The most important and immediate applications of nanonodes are in the biomedical area. Nanonodes can interact with organs and tissues. This is clearly provided due to the nanosize, biocompatibility and biostability. Nanomachines  deployed inside the human body are remotely controlled from the macroscale and over the Internet by an external user such as a healthcare provider as illustrated in Fig. \ref{fig:healthcare}.

Moreover, an immune system can be composed of several nanomachines that protect an organism against diseases. These nanomachines can act in a coordinated way to identify, control, and eliminate foreign and pathogen elements. They could realize tasks of localization and respond to malicious agents and cells, such as cancer cells \cite{patra2013intelligent}, resulting in less aggressive and invasive treatments compared to the existing ones.  

The monitoring of oxygen and cholesterol levels, hormonal disorders, and early diagnosis are some examples of possible applications that can take advantage of intrabody nanosensor networks. The information retrieved by these systems must be accessible outside the body to doctors, nurses, etc. Thus, nanonetworks must provide the proper level of connectivity to deliver the sensed information. In addition, a drug delivery system that is composed of nanonodes could help compensate metabolic diseases such as diabetes. In this scenario, nanosensors and smart glucose reservoirs  can work in a cooperative manner to support regulating mechanisms. One other promising application of the nanosensors is checking for bacteria or viruses in hospitals \cite{balasubramaniam2013realizing}. If contaminating bacteria can be located, it is possible to reduce the number of patients who develop complications such as contagious infections. Finally, manipulation and modification of nanostructures such as molecular sequences and genes can be achieved by nanomachines. The use of nanonetworks will actually allow expanding the potential applications in genetic engineering.

\subsection{Environmental Applications}
Trees, herbs or bushes, release several chemical compositions to the air in order to attract the natural predators of the insects that are attacking them, or to regulate their blooming among different plantations, amongst others \cite{heil2007within}\cite{heil2008long}. Chemical nanosensors \cite{yonzon2005towards} could be used to detect the chemical compounds that are being released and exchanged between plants. Nanonetworks can be build around classical sensor devices which are already deployed in agricultural fields \cite{akyildiz2002wireless}. Other environmental applications include biodiversity control, biodegradation assistance, or air pollution control \cite{riu2006nanosensors}.  
\subsection{Industrial Applications}
\begin{figure}[h!]
\centering
\includegraphics[width=0.65\textwidth]
{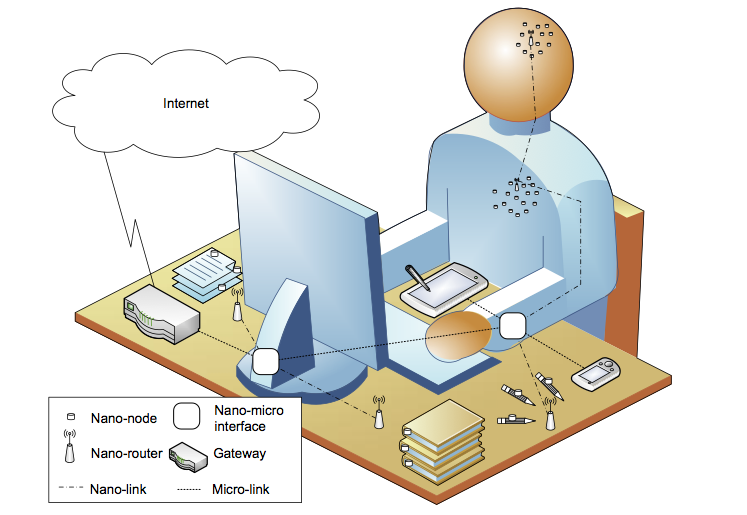}
\caption{Schematic of an intelligent office \cite{akyildiz2010internet}.}
\label{fig:iont}
\end{figure}

The applications of nanotechnology in the development of new industrial and consumer goods range from flexible and stretchable electronic devices \cite{rogers2010materials} to new functionalized nanomaterials for self cleaning anti-microbal textiles \cite{tessier2005antimicrobial}. In addition, the integration of nanomachines with communication capabilities in every single object will allow the interaction  of almost everything in our daily life, from cooking utensils to every element in our working place, or also the components of every device, enabling what we define as the Internet of Nano Things (IoNT) \cite{akyildiz2010internet}. Moreover, as nanocameras and nanophones are developed, in a more futuristic approach, the Internet of Multimedia Nano Things will also become a reality \cite{jornet2012internet}.
Fig. \ref{fig:iont} presents a schematic of a future interconnected office envisaging the concept of nanonetworks where the users can keep track of the location and status of all their belongings in an effortless fashion. 

\section{Current Developments of Nanoscale Devices}

\begin{figure}[h!]
\centering
\includegraphics[width=0.5\textwidth]
{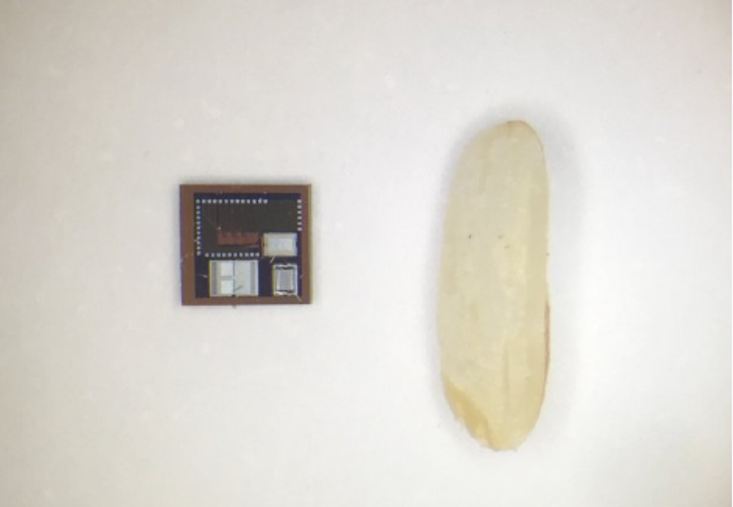}
\caption{Integrated Chip developed by National Applied Research Laboratory with a size comparable to a grain of rice \cite{7063066}.}
\label{fig:rice}
\end{figure}

\begin{figure}[h!]
\centering
\includegraphics[width=0.5\textwidth]
{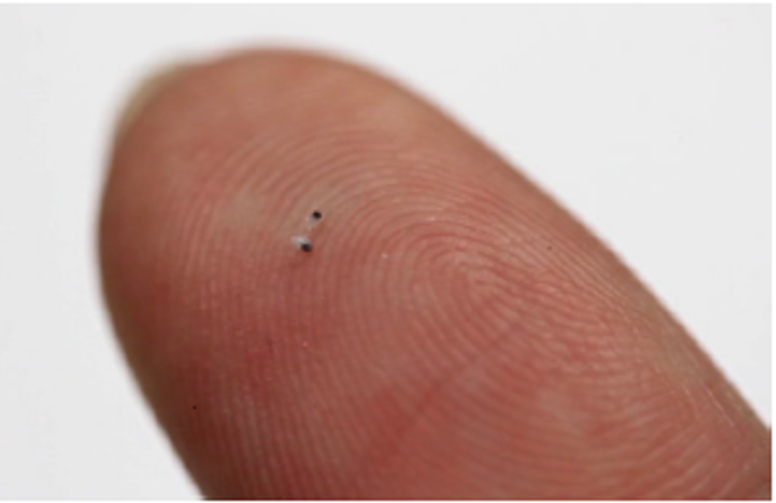}
\caption{Nano-scallop which can swim in bio-fluids \cite{qiu2014swimming}.}
\label{fig:Nanoscallop}
\end{figure}

Due to the developments in micro-fabrication and nanotechnologies, the limits of the  sizes and capabilities of devices have been pushed further. The initial goal of developing small-scale devices is to replace the existing tethered medical devices such as flexible endoscopes and catheters with devices capable of  accessing complex and small regions of the human body like the gastrointestinal tract, spinal cord, and blood capillaries. At the same time, the patient discomfort as well as tissue loss due to sedation would be hugely decreased.   Fig. \ref{fig:rice} demonstrates a chip that is smaller than a grain of rice designed by the National Applied Research Laboratory, using sensor fusion technologies. The micro-robots voyaging around human body were developed recently according to the same principles as well. For example, a tiny permanent magnet, guided inside the human body by a magnetic stereotaxis system was proposed in \cite{meeker1996optimal} while a magnetically driven screw was made to move through tissues as presented in \cite{ishiyama2001swimming}. Micro-mechanical flying insect robots were first created in the University of California, Berkeley \cite{yan2001wing} and then later a solar-powered crawling robot was realized in \cite{hollar2003solar}. The first medical-used capsule endoscopes were applied clinically in 2001 after attaining the  Food and Drug Administration (FDA) approval. Later, the introduction of a crawling mechanism \cite{quirini2008design} and on-board drug delivery mechanism \cite{yim2012design} were marked as  milestones for the development of  capsule endoscopy. A nano-scallop, presented in Fig. \ref{fig:Nanoscallop},  whose size is only a fraction of millimetre and is capable of swimming in biomedical fluids, has been developed at the Max Planck Institute for Intelligent Systems \cite{qiu2014swimming}.
 
On another hand, a surgical nanorobot, programmed or guided by a human surgeon  could act as a semi-autonomous on site surgeon when introduced into the body through  vascular system or body cavities  \cite{patil2008future}. These devices could perform various functions such as diagnosing and correcting lesions by nano-manipulation. Such mechanism is coordinated by an onbaord computer while maintaining contact with the supervising surgeon via coded ultrasound signals. Basically, programmable and controllable microscale robots comprised of nanoscale parts fabricated to nanometer precision will allow medical doctors to execute creative and reconstructive procedures in the human body at the cellular and molecular levels \cite{FreitasJr2005243}.

Besides the research activities on tiny robots, there are also investigations on other applications. A wireless radiation detector was designed to be injected into tumors to detect the level of therapeutic radiation the tumor gets \cite{son2008wireless}. Applying micro-machining techniques, this dosimeter was shrunk to 2 cm long and 2 mm wide in diameter. Overall, reduction in sensor size provides great versatility for incorporation into multiplexed, portable, wearable, as well as implantable medical devices \cite{sagadevan2014recent}. The integration of nanoscale ultrasensitive biosensors with other medical instruments will open the door to emerging medical fields, including point-of-care diagnostics and ubiquitous healthcare systems \cite{touhami2014biosensors}. Specifically, the future impact of nanobiosensor systems for point-of-care diagnostics will be unmatched. This technology will revolutionize conventional medical practices by enabling early diagnosis of chronic debilitating diseases, ultrasensitive detection of pathogens, and long-term monitoring of patients using biocompatible integrated medical instrumentation \cite{santrachcurrent}.

\newpage 
\bibliographystyle{IEEEtran}
\bibliography{main}

\begin{thebibliography}{10}
\providecommand{\url}[1]{#1}
\csname url@samestyle\endcsname
\providecommand{\newblock}{\relax}
\providecommand{\bibinfo}[2]{#2}
\providecommand{\BIBentrySTDinterwordspacing}{\spaceskip=0pt\relax}
\providecommand{\BIBentryALTinterwordstretchfactor}{4}
\providecommand{\BIBentryALTinterwordspacing}{\spaceskip=\fontdimen2\font plus
\BIBentryALTinterwordstretchfactor\fontdimen3\font minus
  \fontdimen4\font\relax}
\providecommand{\BIBforeignlanguage}[2]{{%
\expandafter\ifx\csname l@#1\endcsname\relax
\typeout{** WARNING: IEEEtran.bst: No hyphenation pattern has been}%
\typeout{** loaded for the language `#1'. Using the pattern for}%
\typeout{** the default language instead.}%
\else
\language=\csname l@#1\endcsname
\fi
#2}}
\providecommand{\BIBdecl}{\relax}
\BIBdecl

\bibitem{omar2016uwb}
A.~Omar and R.~Shubair, ``Uwb coplanar waveguide-fed-coplanar strips spiral
  antenna,'' in \emph{2016 10th European Conference on Antennas and Propagation
  (EuCAP)}.\hskip 1em plus 0.5em minus 0.4em\relax IEEE, 2016, pp. 1--2.

\bibitem{elayan2017terahertz}
H.~Elayan, R.~M. Shubair, J.~M. Jornet, and P.~Johari, ``Terahertz channel
  model and link budget analysis for intrabody nanoscale communication,''
  \emph{IEEE transactions on nanobioscience}, vol.~16, no.~6, pp. 491--503,
  2017.

\bibitem{elayan2017wireless}
H.~Elayan, R.~M. Shubair, and A.~Kiourti, ``Wireless sensors for medical
  applications: Current status and future challenges,'' in \emph{2017 11th
  European Conference on Antennas and Propagation (EUCAP)}.\hskip 1em plus
  0.5em minus 0.4em\relax IEEE, 2017, pp. 2478--2482.

\bibitem{elayan2016channel}
H.~Elayan and R.~M. Shubair, ``On channel characterization in human body
  communication for medical monitoring systems,'' in \emph{2016 17th
  International Symposium on Antenna Technology and Applied Electromagnetics
  (ANTEM)}.\hskip 1em plus 0.5em minus 0.4em\relax IEEE, 2016, pp. 1--2.

\bibitem{elayan2016vivo}
H.~Elayan, R.~M. Shubair, A.~Alomainy, and K.~Yang, ``In-vivo terahertz em
  channel characterization for nano-communications in wbans,'' in \emph{2016
  IEEE International Symposium on Antennas and Propagation (APSURSI)}.\hskip
  1em plus 0.5em minus 0.4em\relax IEEE, 2016, pp. 979--980.

\bibitem{elayan2017bio}
H.~Elayan, R.~M. Shubair, and J.~M. Jornet, ``Bio-electromagnetic thz
  propagation modeling for in-vivo wireless nanosensor networks,'' in
  \emph{2017 11th European Conference on Antennas and Propagation
  (EuCAP)}.\hskip 1em plus 0.5em minus 0.4em\relax IEEE, 2017, pp. 426--430.

\bibitem{elayan2018end}
H.~Elayan, C.~Stefanini, R.~M. Shubair, and J.~M. Jornet, ``End-to-end noise
  model for intra-body terahertz nanoscale communication,'' \emph{IEEE
  transactions on nanobioscience}, vol.~17, no.~4, pp. 464--473, 2018.

\bibitem{elayan2017photothermal}
H.~Elayan, P.~Johari, R.~M. Shubair, and J.~M. Jornet, ``Photothermal modeling
  and analysis of intrabody terahertz nanoscale communication,'' \emph{IEEE
  transactions on nanobioscience}, vol.~16, no.~8, pp. 755--763, 2017.

\bibitem{elayan2017multi}
H.~Elayan, R.~M. Shubair, J.~M. Jornet, and R.~Mittra, ``Multi-layer intrabody
  terahertz wave propagation model for nanobiosensing applications,''
  \emph{Nano communication networks}, vol.~14, pp. 9--15, 2017.

\bibitem{elayan2018vivo}
H.~Elayan, R.~M. Shubair, and N.~Almoosa, ``In vivo communication in wireless
  body area networks,'' in \emph{Information Innovation Technology in Smart
  Cities}.\hskip 1em plus 0.5em minus 0.4em\relax Springer, 2018, pp. 273--287.

\bibitem{alnabooda2017terahertz}
M.~O. AlNabooda, R.~M. Shubair, N.~R. Rishani, and G.~Aldabbagh, ``Terahertz
  spectroscopy and imaging for the detection and identification of illicit
  drugs,'' in \emph{2017 Sensors networks smart and emerging technologies
  (SENSET)}.\hskip 1em plus 0.5em minus 0.4em\relax IEEE, 2017, pp. 1--4.

\bibitem{khan2017ultra}
M.~S. Khan, A.-D. Capobianco, A.~Iftikhar, R.~M. Shubair, D.~E. Anagnostou, and
  B.~D. Braaten, ``Ultra-compact dual-polarised uwb mimo antenna with meandered
  feeding lines,'' \emph{IET Microwaves, Antennas \& Propagation}, vol.~11,
  no.~7, pp. 997--1002, 2017.

\bibitem{khan2017compact}
M.~S. Khan, A.-D. Capobianco, S.~M. Asif, D.~E. Anagnostou, R.~M. Shubair, and
  B.~D. Braaten, ``A compact csrr-enabled uwb diversity antenna,'' \emph{IEEE
  Antennas and Wireless Propagation Letters}, vol.~16, pp. 808--812, 2017.

\bibitem{shubair2015novel}
R.~M. Shubair, A.~M. AlShamsi, K.~Khalaf, and A.~Kiourti, ``Novel miniature
  wearable microstrip antennas for ism-band biomedical telemetry,'' in
  \emph{2015 Loughborough Antennas \& Propagation Conference (LAPC)}.\hskip 1em
  plus 0.5em minus 0.4em\relax IEEE, 2015, pp. 1--4.

\bibitem{shubair2015vivo}
R.~M. Shubair and H.~Elayan, ``In vivo wireless body communications:
  State-of-the-art and future directions,'' in \emph{2015 Loughborough Antennas
  \& Propagation Conference (LAPC)}.\hskip 1em plus 0.5em minus 0.4em\relax
  IEEE, 2015, pp. 1--5.

\bibitem{alayyan2009mmse}
F.~O. Alayyan, R.~M. Shubair, Y.~H. Leung, A.~M. Zoubir, and O.~Alketbi, ``On
  mmse methods for blind identification of ofdm-based simo systems,'' in
  \emph{2009 IFIP International Conference on Wireless and Optical
  Communications Networks}.\hskip 1em plus 0.5em minus 0.4em\relax IEEE, 2009,
  pp. 1--5.

\bibitem{elayan2018towards}
H.~Elayan and R.~M. Shubair, ``Towards an intelligent deployment of wireless
  sensor networks,'' in \emph{Information Innovation Technology in Smart
  Cities}.\hskip 1em plus 0.5em minus 0.4em\relax Springer, 2018, pp. 235--250.

\bibitem{kulaib2015improved}
A.~Kulaib, R.~Shubair, M.~Al-Qutayri, and J.~W. Ng, ``Improved dv-hop
  localization using node repositioning and clustering,'' in \emph{2015
  International Conference on Communications, Signal Processing, and their
  Applications (ICCSPA'15)}.\hskip 1em plus 0.5em minus 0.4em\relax IEEE, 2015,
  pp. 1--6.

\bibitem{elayan2018stochastic}
H.~Elayan, C.~Stefanini, R.~M. Shubair, and J.~M. Jornet, ``Stochastic noise
  model for intra-body terahertz nanoscale communication,'' in
  \emph{Proceedings of the 5th ACM International Conference on Nanoscale
  Computing and Communication}.\hskip 1em plus 0.5em minus 0.4em\relax ACM,
  2018, p.~8.

\bibitem{elayan2018graphene}
H.~Elayan, R.~M. Shubair, J.~M. Jornet, A.~Kiourti, and R.~Mittra,
  ``Graphene-based spiral nanoantenna for intrabody communication at
  terahertz,'' in \emph{2018 IEEE International Symposium on Antennas and
  Propagation \& USNC/URSI National Radio Science Meeting}.\hskip 1em plus
  0.5em minus 0.4em\relax IEEE, 2018, pp. 799--800.

\bibitem{elayan2018characterising}
H.~Elayan, R.~M. Shubair, and J.~M. Jornet, ``Characterising thz propagation
  and intrabody thermal absorption in iwnsns,'' \emph{IET Microwaves, Antennas
  \& Propagation}, vol.~12, no.~4, pp. 525--532, 2018.

\bibitem{elayan2018terahertz}
H.~Elayan, O.~Amin, R.~M. Shubair, and M.-S. Alouini, ``Terahertz
  communication: The opportunities of wireless technology beyond 5g,'' in
  \emph{2018 International Conference on Advanced Communication Technologies
  and Networking (CommNet)}.\hskip 1em plus 0.5em minus 0.4em\relax IEEE, 2018,
  pp. 1--5.

\bibitem{alharbi2018flexible}
S.~Alharbi, R.~M. Shubair, and A.~Kiourti, ``Flexible antennas for wearable
  applications: Recent advances and design challenges,'' 2018.

\bibitem{kiourti2017implantable}
A.~Kiourti and R.~M. Shubair, ``Implantable and ingestible sensors for wireless
  physiological monitoring: A review,'' in \emph{2017 IEEE International
  Symposium on Antennas and Propagation \& USNC/URSI National Radio Science
  Meeting}.\hskip 1em plus 0.5em minus 0.4em\relax IEEE, 2017, pp. 1677--1678.

\bibitem{khan2017second}
O.~M. Khan, R.~M. Shubair, and Q.~U. Islam, ``Second order flamenco fractal
  antenna for industrial scientific and medical applications,'' in \emph{2017
  International Conference on Electrical and Computing Technologies and
  Applications (ICECTA)}.\hskip 1em plus 0.5em minus 0.4em\relax IEEE, 2017,
  pp. 1--3.

\bibitem{ibrahim2017compact}
A.~A. Ibrahim, J.~Machac, and R.~M. Shubair, ``Compact uwb mimo antenna with
  pattern diversity and band rejection characteristics,'' \emph{Microwave and
  Optical Technology Letters}, vol.~59, no.~6, pp. 1460--1464, 2017.

\bibitem{khan2016properties}
M.~S. Khan, A.-D. Capobianco, S.~M. Asif, A.~Iftikhar, B.~D. Braaten, and R.~M.
  Shubair, ``A properties comparison between copper and graphene-based uwb mimo
  planar antennas,'' in \emph{2016 IEEE International Symposium on Antennas and
  Propagation (APSURSI)}.\hskip 1em plus 0.5em minus 0.4em\relax IEEE, 2016,
  pp. 1767--1768.

\bibitem{khan2016pattern}
------, ``A pattern reconfigurable printed patch antenna,'' in \emph{2016 IEEE
  International Symposium on Antennas and Propagation (APSURSI)}.\hskip 1em
  plus 0.5em minus 0.4em\relax IEEE, 2016, pp. 2149--2150.

\bibitem{elayan2016revolutionizing}
H.~Elayan, R.~M. Shubair, and N.~Almoosa, ``Revolutionizing the healthcare of
  the future through nanomedicine: Opportunities and challenges,'' in
  \emph{2016 12th International Conference on Innovations in Information
  Technology (IIT)}.\hskip 1em plus 0.5em minus 0.4em\relax IEEE, 2016, pp.
  1--5.

\bibitem{el2016design}
M.~El~Shorbagy, R.~M. Shubair, M.~I. AlHajri, and N.~K. Mallat, ``On the design
  of millimetre-wave antennas for 5g,'' in \emph{2016 16th Mediterranean
  Microwave Symposium (MMS)}.\hskip 1em plus 0.5em minus 0.4em\relax IEEE,
  2016, pp. 1--4.

\bibitem{bazazeh2016biomarker}
D.~Bazazeh, R.~M. Shubair, and W.~Q. Malik, ``Biomarker discovery and
  validation for parkinson's disease: A machine learning approach,'' in
  \emph{2016 International Conference on Bio-engineering for Smart Technologies
  (BioSMART)}.\hskip 1em plus 0.5em minus 0.4em\relax IEEE, 2016, pp. 1--6.

\bibitem{bazazeh2016comparative}
D.~Bazazeh and R.~Shubair, ``Comparative study of machine learning algorithms
  for breast cancer detection and diagnosis,'' in \emph{2016 5th International
  Conference on Electronic Devices, Systems and Applications (ICEDSA)}.\hskip
  1em plus 0.5em minus 0.4em\relax IEEE, 2016, pp. 1--4.

\bibitem{albreiki2016coding}
S.~Albreiki, A.~AlAli, and R.~M. Shubair, ``Coding brain neurons via electrical
  network models for neuro-signal synthesis in computational neuroscience,'' in
  \emph{2016 5th international conference on electronic devices, systems and
  applications (ICEDSA)}.\hskip 1em plus 0.5em minus 0.4em\relax IEEE, 2016,
  pp. 1--5.

\bibitem{shubair2015survey}
R.~M. Shubair and H.~Elayan, ``A survey of in vivo wban communications and
  networking: Research issues and challenges,'' in \emph{2015 11th
  International Conference on Innovations in Information Technology
  (IIT)}.\hskip 1em plus 0.5em minus 0.4em\relax IEEE, 2015, pp. 11--16.

\bibitem{elsalamouny2015novel}
M.~Y. ElSalamouny and R.~M. Shubair, ``Novel design of compact low-profile
  multi-band microstrip antennas for medical applications,'' in \emph{2015
  Loughborough Antennas \& Propagation Conference (LAPC)}.\hskip 1em plus 0.5em
  minus 0.4em\relax IEEE, 2015, pp. 1--4.

\bibitem{nwalozie2013simple}
G.~Nwalozie, V.~Okorogu, S.~Maduadichie, and A.~Adenola, ``A simple comparative
  evaluation of adaptive beam forming algorithms,'' \emph{International Journal
  of Engineering and Innovative Technology (IJEIT)}, vol.~2, no.~7, 2013.

\bibitem{che2008propagation}
W.~Che, C.~Li, P.~Russer, and Y.~Chow, ``Propagation and band broadening effect
  of planar integrated ridged waveguide in multilayer dielectric substrates,''
  in \emph{2008 IEEE MTT-S International Microwave Symposium Digest}.\hskip 1em
  plus 0.5em minus 0.4em\relax IEEE, 2008, pp. 217--220.

\bibitem{bakhar2009eigen}
M.~Bakhar and D.~P. Hunagund, ``Eigen structure based direction of arrival
  estimation algorithms for smart antenna systems,'' \emph{IJCSNS International
  Journal of Computer Science and Network Security}, vol.~9, no.~11, pp.
  96--100, 2009.

\bibitem{khan2018compact}
M.~Khan, F.~Rigobello, B.~Ijaz, E.~Autizi, A.~Capobianco, R.~Shubair, and
  S.~Khan, ``Compact 3-d eight elements uwb-mimo array,'' \emph{Microwave and
  Optical Technology Letters}, vol.~60, no.~8, pp. 1967--1971, 2018.

\bibitem{khan2016compact}
M.~S. Khan, A.-D. Capobianco, A.~Iftikhar, S.~Asif, and B.~D. Braaten, ``A
  compact dual polarized ultrawideband multiple-input-multiple-output
  antenna,'' \emph{Microwave and Optical Technology Letters}, vol.~58, no.~1,
  pp. 163--166, 2016.

\bibitem{al2005direction}
M.~Al-Nuaimi, R.~Shubair, and K.~Al-Midfa, ``Direction of arrival estimation in
  wireless mobile communications using minimum variance distortionless
  response,'' in \emph{The Second International Conference on Innovations in
  Information Technology (IIT’05)}, 2005, pp. 1--5.

\bibitem{akyildiz2008nanonetworks}
I.~F. Akyildiz, F.~Brunetti, and C.~Bl{\'a}zquez, ``Nanonetworks: A new
  communication paradigm,'' \emph{Computer Networks}, vol.~52, no.~12, pp.
  2260--2279, 2008.

\bibitem{akyildiz2010electromagnetic}
I.~F. Akyildiz and J.~M. Jornet, ``Electromagnetic wireless nanosensor
  networks,'' \emph{Nano Communication Networks}, vol.~1, no.~1, pp. 3--19,
  2010.

\bibitem{rutherglen2009nanoelectromagnetics}
C.~Rutherglen and P.~Burke, ``Nanoelectromagnetics: circuit and electromagnetic
  properties of carbon nanotubes,'' \emph{small}, vol.~5, no.~8, pp. 884--906,
  2009.

\bibitem{akyildiz2010propagation}
I.~F. Akyildiz, J.~M. Jornet, and M.~Pierobon, ``Propagation models for
  nanocommunication networks,'' in \emph{Proceedings of the Fourth European
  Conference on Antennas and Propagation}.\hskip 1em plus 0.5em minus
  0.4em\relax IEEE, 2010, pp. 1--5.

\bibitem{jornet2013fundamentals}
J.~M. Jornet and I.~F. Akyildiz, ``Fundamentals of electromagnetic nanonetworks
  in the terahertz band,'' \emph{Foundations and Trends{\textregistered} in
  Networking}, vol.~7, no. 2-3, pp. 77--233, 2013.

\bibitem{ponomarenko2008chaotic}
L.~Ponomarenko, F.~Schedin, M.~Katsnelson, R.~Yang, E.~Hill, K.~Novoselov, and
  A.~Geim, ``Chaotic dirac billiard in graphene quantum dots,'' \emph{Science},
  vol. 320, no. 5874, pp. 356--358, 2008.

\bibitem{canovas2016conceptual}
S.~Canovas-Carrasco, A.-J. Garcia-Sanchez, F.~Garcia-Sanchez, and
  J.~Garcia-Haro, ``Conceptual design of a nano-networking device,''
  \emph{Sensors}, vol.~16, no.~12, p. 2104, 2016.

\bibitem{bennewitz2002atomic}
R.~Bennewitz, J.~N. Crain, A.~Kirakosian, J.~Lin, J.~McChesney, D.~Petrovykh,
  and F.~Himpsel, ``Atomic scale memory at a silicon surface,''
  \emph{Nanotechnology}, vol.~13, no.~4, p. 499, 2002.

\bibitem{parkin2008magnetic}
S.~S. Parkin, M.~Hayashi, and L.~Thomas, ``Magnetic domain-wall racetrack
  memory,'' \emph{Science}, vol. 320, no. 5873, pp. 190--194, 2008.

\bibitem{ji2011multilayer}
L.~Ji, Z.~Tan, T.~Kuykendall, E.~J. An, Y.~Fu, V.~Battaglia, and Y.~Zhang,
  ``Multilayer nanoassembly of sn-nanopillar arrays sandwiched between graphene
  layers for high-capacity lithium storage,'' \emph{Energy \& Environmental
  Science}, vol.~4, no.~9, pp. 3611--3616, 2011.

\bibitem{wang2008towards}
Z.~L. Wang, ``Towards self-powered nanosystems: from nanogenerators to
  nanopiezotronics,'' \emph{Advanced Functional Materials}, vol.~18, no.~22,
  pp. 3553--3567, 2008.

\bibitem{wang2006piezoelectric}
Z.~L. Wang and J.~Song, ``Piezoelectric nanogenerators based on zinc oxide
  nanowire arrays,'' \emph{Science}, vol. 312, no. 5771, pp. 242--246, 2006.

\bibitem{hierold2007nano}
C.~Hierold, A.~Jungen, C.~Stampfer, and T.~Helbling, ``Nano electromechanical
  sensors based on carbon nanotubes,'' \emph{Sensors and Actuators A:
  Physical}, vol. 136, no.~1, pp. 51--61, 2007.

\bibitem{akyildiz2010internet}
I.~F. Akyildiz and J.~M. Jornet, ``The internet of nano-things,'' \emph{IEEE
  Wireless Communications}, vol.~17, no.~6, pp. 58--63, 2010.

\bibitem{patra2013intelligent}
D.~Patra, S.~Sengupta, W.~Duan, H.~Zhang, R.~Pavlick, and A.~Sen,
  ``Intelligent, self-powered, drug delivery systems,'' \emph{Nanoscale},
  vol.~5, no.~4, pp. 1273--1283, 2013.

\bibitem{balasubramaniam2013realizing}
S.~Balasubramaniam and J.~Kangasharju, ``Realizing the internet of nano things:
  challenges, solutions, and applications,'' \emph{Computer}, vol.~46, no.~2,
  pp. 62--68, 2013.

\bibitem{heil2007within}
M.~Heil and J.~C.~S. Bueno, ``Within-plant signaling by volatiles leads to
  induction and priming of an indirect plant defense in nature,''
  \emph{Proceedings of the National Academy of Sciences}, vol. 104, no.~13, pp.
  5467--5472, 2007.

\bibitem{heil2008long}
M.~Heil and J.~Ton, ``Long-distance signalling in plant defence,'' \emph{Trends
  in plant science}, vol.~13, no.~6, pp. 264--272, 2008.

\bibitem{yonzon2005towards}
C.~R. Yonzon, D.~A. Stuart, X.~Zhang, A.~D. McFarland, C.~L. Haynes, and R.~P.
  Van~Duyne, ``Towards advanced chemical and biological nanosensorsan
  overview,'' \emph{Talanta}, vol.~67, no.~3, pp. 438--448, 2005.

\bibitem{akyildiz2002wireless}
I.~F. Akyildiz, W.~Su, Y.~Sankarasubramaniam, and E.~Cayirci, ``Wireless sensor
  networks: a survey,'' \emph{Computer networks}, vol.~38, no.~4, pp. 393--422,
  2002.

\bibitem{riu2006nanosensors}
J.~Riu, A.~Maroto, and F.~X. Rius, ``Nanosensors in environmental analysis,''
  \emph{Talanta}, vol.~69, no.~2, pp. 288--301, 2006.

\bibitem{rogers2010materials}
J.~A. Rogers, T.~Someya, and Y.~Huang, ``Materials and mechanics for
  stretchable electronics,'' \emph{Science}, vol. 327, no. 5973, pp.
  1603--1607, 2010.

\bibitem{tessier2005antimicrobial}
D.~Tessier, I.~Radu, and M.~Filteau, ``Antimicrobial fabrics coated with
  nano-sized silver salt crystals,'' in \emph{NSTI Nanotech}, vol.~1, 2005, pp.
  762--764.

\bibitem{jornet2012internet}
J.~M. Jornet and I.~F. Akyildiz, ``The internet of multimedia nano-things,''
  \emph{Nano Communication Networks}, vol.~3, no.~4, pp. 242--251, 2012.

\bibitem{meeker1996optimal}
D.~C. Meeker, E.~H. Maslen, R.~C. Ritter, and F.~M. Creighton, ``Optimal
  realization of arbitrary forces in a magnetic stereotaxis system,''
  \emph{IEEE Transactions on Magnetics}, vol.~32, no.~2, pp. 320--328, 1996.

\bibitem{ishiyama2001swimming}
K.~Ishiyama, M.~Sendoh, A.~Yamazaki, and K.~Arai, ``Swimming micro-machine
  driven by magnetic torque,'' \emph{Sensors and Actuators A: Physical},
  vol.~91, no.~1, pp. 141--144, 2001.

\bibitem{yan2001wing}
J.~Yan, S.~Avadhanula, J.~Birch, M.~Dickinson, M.~Sitti, T.~Su, and R.~Fearing,
  ``Wing transmission for a micromechanical flying insect,'' \emph{Journal of
  Micromechatronics}, vol.~1, no.~3, pp. 221--237, 2001.

\bibitem{hollar2003solar}
S.~Hollar, A.~Flynn, C.~Bellew, and K.~Pister, ``Solar powered 10 mg silicon
  robot,'' in \emph{Micro Electro Mechanical Systems, 2003. MEMS-03 Kyoto. IEEE
  The Sixteenth Annual International Conference on}.\hskip 1em plus 0.5em minus
  0.4em\relax IEEE, 2003, pp. 706--711.

\bibitem{quirini2008design}
M.~Quirini, A.~Menciassi, S.~Scapellato, C.~Stefanini, and P.~Dario, ``Design
  and fabrication of a motor legged capsule for the active exploration of the
  gastrointestinal tract,'' \emph{IEEE/ASME transactions on mechatronics},
  vol.~13, no.~2, pp. 169--179, 2008.

\bibitem{yim2012design}
S.~Yim and M.~Sitti, ``Design and rolling locomotion of a magnetically actuated
  soft capsule endoscope,'' \emph{IEEE Transactions on Robotics}, vol.~28,
  no.~1, pp. 183--194, 2012.

\bibitem{qiu2014swimming}
T.~Qiu, T.-C. Lee, A.~G. Mark, K.~I. Morozov, R.~M{\"u}nster, O.~Mierka,
  S.~Turek, A.~M. Leshansky, and P.~Fischer, ``Swimming by reciprocal motion at
  low reynolds number,'' \emph{Nature communications}, vol.~5, 2014.

\bibitem{7063066}
J.~Zhou, T.~H. Chuang, T.~Dinc, and H.~Krishnaswamy, ``Reconfigurable receiver
  with >20mhz bandwidth self-interference cancellation suitable for fdd,
  co-existence and full-duplex applications,'' in \emph{2015 IEEE International
  Solid-State Circuits Conference - (ISSCC) Digest of Technical Papers}, Feb
  2015, pp. 1--3.

\bibitem{patil2008future}
M.~Patil, D.~S. Mehta, S.~Guvva \emph{et~al.}, ``Future impact of
  nanotechnology on medicine and dentistry,'' \emph{Journal of Indian society
  of periodontology}, vol.~12, no.~2, p.~34, 2008.

\bibitem{FreitasJr2005243}
\BIBentryALTinterwordspacing
R.~A.~F. Jr., ``Nanotechnology, nanomedicine and nanosurgery,''
  \emph{International Journal of Surgery}, vol.~3, no.~4, pp. 243 -- 246, 2005.
  [Online]. Available:
  \url{http://www.sciencedirect.com/science/article/pii/S1743919105001299}
\BIBentrySTDinterwordspacing

\bibitem{son2008wireless}
C.~Son and B.~Ziaie, ``A wireless implantable passive microdosimeter for
  radiation oncology,'' \emph{IEEE Transactions on Biomedical Engineering},
  vol.~55, no.~6, pp. 1772--1775, 2008.

\bibitem{sagadevan2014recent}
S.~Sagadevan and M.~Periasamy, ``Recent trends in nanobiosensors and their
  applications-a review,'' \emph{Rev Adv Mater Sci}, vol.~36, pp. 62--69, 2014.

\bibitem{touhami2014biosensors}
A.~Touhami, ``Biosensors and nanobiosensors: Design and applications.''

\bibitem{santrachcurrent}
P.~J. Santrach, ``Current \& future applications of point of care testing,''
  \emph{Mayo Clinic, www. cdc. gov Search PubMed}.

\end{thebibliography}

\end{document}